\documentclass[twocolumn,prb,aps,superscriptaddress]{revtex4-1}

\usepackage[T1]{fontenc}
\usepackage[latin9]{inputenc}

\usepackage{amsmath}
\usepackage{amssymb}
\usepackage{braket}
\usepackage{xcolor}
\usepackage[normalem]{ulem}
\usepackage{longtable}

\numberwithin{equation}{section}

\allowdisplaybreaks

\usepackage{graphicx}
\usepackage[colorlinks=true]{hyperref}  
\hypersetup{
    bookmarks=true,         
    unicode=false,          
    pdftoolbar=true,        
    pdfmenubar=true,        
    pdffitwindow=false,     
    pdfstartview={FitH},    
    pdftitle={Band Engineering of Dirac cones in Iron Chalcogenides},    
    pdfauthor={Lauke, Heid, Merz, Haghighirad, Schmalian},     
    pdfsubject={},   
    pdfcreator={},   
    pdfproducer={}, 
    pdfkeywords={} {} {}, 
    pdfnewwindow=true,      
    colorlinks=true,       
    linkcolor=blue, 
    citecolor=blue,        
    filecolor=magenta,      
    urlcolor=blue           
}

\newcommand{\equref}[1]{Eq.~(\ref{#1})}

\newcommand{\secref}[1]{Sec.~\ref{#1}}

\newcommand{\figref}[1]{Fig.~\ref{#1}}
\newcommand{\refcite}[1]{Ref.~\onlinecite{#1}}
\newcommand{\refscite}[1]{Refs.~\onlinecite{#1}}

\newcommand{\tableref}[1]{Table~\ref{#1}}
\newcommand{\appref}[1]{Appendix~\ref{#1}}

\renewcommand{\Im}{\text{Im}}
\renewcommand{\vec}[1]{\boldsymbol{#1}}
\newcommand{\mbeq}{\overset{!}{=}}

\begin{document}

\title{ Band Engineering of Dirac cones in Iron Chalcogenides}
\author{Lars Lauke}
\affiliation{Institute for Quantum Materials and Technologies, Karlsruhe Institute of Technology (KIT), 76131 Karlsruhe, Germany}
\author{Rolf Heid}
\affiliation{Institute for Quantum Materials and Technologies, Karlsruhe Institute of Technology (KIT), 76131 Karlsruhe, Germany}
\author{Michael Merz}
\affiliation{Institute for Quantum Materials and Technologies, Karlsruhe Institute of Technology (KIT), 76131 Karlsruhe, Germany}
\author{Thomas Wolf}
\affiliation{Institute for Quantum Materials and Technologies, Karlsruhe Institute of Technology (KIT), 76131 Karlsruhe, Germany}
\author{Amir-Abbas Haghighirad}
\affiliation{Institute for Quantum Materials and Technologies, Karlsruhe Institute of Technology (KIT), 76131 Karlsruhe, Germany}
\author{J\"org Schmalian}
\affiliation{Institute for Theory of Condensed Matter, Karlsruhe Institute of Technology (KIT), 76131 Karlsruhe, Germany}
\affiliation{Institute for Quantum Materials and Technologies, Karlsruhe Institute of Technology (KIT), 76131 Karlsruhe, Germany}

\begin{abstract}
By band engineering the iron chalcogenide Fe(Se,Te) via ab-initio calculations, we search for topological surface states and realizations of Majorana bound states. Proposed topological states are expected to occur for non-stoichiometric compositions on a surface Dirac cone where issues like disorder scattering and charge transfer between relevant electronic states have to be addressed. However, this surface Dirac cone is well above the Fermi-level. Our goal is to theoretically design a substituted crystal in which the surface Dirac cone is shifted towards the Fermi-level by modifying the bulk material without disturbing the surface.  Going beyond conventional density functional theory (DFT), we apply the coherent potential approximation (BEB-CPA) in a mixed basis pseudo-potential framework to scan the substitutional phase-space of co-substitutions on the Se-sites. We have identified iodine as a promising candidate for intrinsic doping. Our specific proposal  is that FeSe$_{0.325}$I$_{0.175}$Te$_{0.5}$  is a very likely candidate to exhibit a Dirac cone right at the Fermi energy without inducing strong disorder scattering.  
\end{abstract}
\maketitle

\section{Introduction}
\label{sec:intro}
In recent years the search for solid state systems that host topologically protected surface states have attracted significant attention. In addition to topological insulators, topological superconductors are promising given the interesting properties of the corresponding Majorana bound states\cite{Kitaev_2001,RevModPhys.80.1083,Alicea_2012,Sato_2017}. One avenue towards topological superconductivity is by creating hetero-structures of semiconductors and conventional superconductors, the latter with rather low transition temperatures. A material with high superconducting transition temperature and intrinsic topological superconductivity is clearly desirable.  A very promising systems for a high-$T_{c}$, single crystal realization of Majorana bound states is the Fe-based superconductor FeSe$_{1-x}$Te$_{x}$. Its transition temperature can be brought up to $~30\, {\rm K}$ under external pressure\cite{JPS_Tc} and even above $40\, {\rm K}$ in monolayer thin films\cite{PhysRevB.91.220503}. Futhermore, superconductivity has been observed for a wide range of composition $x$\cite{PhysRevB.78.224503,PhysRevB.79.094521,Yeh_2008}. In addition to the simple structure (see \figref{fig:orb}(a)), it exhibits a high tunability of its internal parameters by chemical substitution\cite{PhysRevB.79.054503,TopCharacter,PhysRevB.92.035104}.
Most notably, FeSe$_{0.5}$Te$_{0.5}$ was argued to possess a non-trivial band topology characterized by a $\mathbb{Z}_{2}$ topological index, hosting a surface Dirac cone. In combination with the proximity to bulk superconductivity\cite{PhysRevLett.100.096407}, this could lead to Majorana bound states\cite{ZhangPRB,TopCharacter,Wang333}. 

However, the main difficulty is the location of said Dirac cone well above the Fermi-level, rendering it irrelevant with respect to experiment. 
Recent attempts to circumvent this problem via surface deposition have yielded promising results\cite{ZhangPRB}. A major drawback of this strategy is the inevitable distortion of surface transport, which is of great interest for systems with topologically protected surface states. 
The main goal of this paper is to provide a strategy to bring the surface Dirac cone closer to the Fermi-level, and thus making the surface states experimentally accessible, while preserving surface transport. 

To this end, we investigate \textit{intrinsic} doping by employing the coherent potential approximation (CPA) to virtually design an appropriate crystal of the form FeSe$_{1-x-y}$Te$_{x}$\textit{A}$_{y}$, where \textit{A} denotes a generic substitution of concentration $y$. The specific strategy of our band-structure engineering is to modify the location of electronic states of  $p_z$-character  by chalcogen, i.e., by Te-substitution and combine this with substitutions that change the intrinsic doping without causing strong impurity scattering. The most promising candidate is a substitution of selenium by a modest amount of iodine. Other approaches such as substitutions of iron by other transition metals may also affect the doping but introduce too strong impurity scattering. The specific proposal therefore is that FeSe$_{0.325}$I$_{0.175}$Te$_{0.5}$  is a very likely candidate to exhibit a Dirac cone right at the Fermi energy.
For our approach to be quantitatively reliable, experimentally obtained structural parameters are essential. The lattice parameters and atomic positions used in our electronic structure calculations were obtained from refined x-ray diffraction data.  
We begin with a brief introduction of the CPA and discuss the effects of Te-substitution and intrinsic doping on the basis of our bandstructure calculations in \secref{sec:Results}, followed by our conclusion in \secref{sec:Conclusion}.

\section{Model and formalism}
\label{ModelIntro}
The goal of this paper is to obtain quantitative first-principles based insight into the electronic structure of substitutionally disordered systems. Our results  are obtained within the {\em ab initio} version of the coherent potential approximation (CPA) due to Blackman, Esterling and Berk (BEB)\cite{BEB}. For convenience we summarize the main idea of the CPA and of the BEB-version of it in the appendix. A key advantage of this formalism is that it offers a feasible treatment of realistic compounds with substitutional disorder that goes beyond the scope of simplified model Hamiltonians. Chemical species-dependent  hopping- and onsite matrix elements  are extracted from \textit{ab initio} DFT calculations.  Our approach also builds on the treatment by Koepernik \textit{et al} who extended the BEB formalism to include multiple orbital degrees of freedom per site and chemical species\cite{Koepernik}. In addition we build our work on the implementation by Herbig \textit{et al}\cite{Herbig1}. As input for our CPA calculations we use DFT results obtained from the mixed-basis pseudo-potential program (MBPP) developed by Meyer \textit{et al.} \cite{MBPP}.

As we are interested in local quantities, we rely on a LCAO-description of the orbitals, where
\begin{equation*}
	\phi_{i\mu}^{P}(\mathbf{r})=\phi_{\mu}^{P}(\mathbf{r}-\mathbf{R}_{i})=\langle \mathbf{r}|iP\mu\rangle,
\end{equation*}
with site index $i$ of an atom of species $P$ located at position $\mathbf{R}_{i}$. Here, $\mu=(l,m)$ is a combined orbital index with orbital angular momentum $l$ and magnetic quantum number $m$. Furthermore, the orbitals are expressed in real spherical harmonics $K_{lm}(\hat{r})$,
\begin{equation*}
	\phi_{\mu}^{P}(\mathbf{r})=\phi_{lm}^{P}(\mathbf{r})=i^l f_{l}^{P}(r)K_{lm}(\hat{r}),
\end{equation*}
where $f_l^P$ are radial and species-dependent functions with $r=|\mathbf{r}|$ and $K_{lm}$ depend only on the angle via $\hat{r}=\mathbf{r}/r=(\vartheta,\varphi)$. Being a non-orthonormal basis set, the local orbitals have a non-vanishing overlap,
\begin{equation*}
	\underline{S}_{i\mu,j\nu}^{P,Q} = \langle iP\mu|jQ\nu\rangle = \int d^3r \big[\phi_{i\mu}^{P}(\mathbf{r})\big]^* \phi_{j\nu}^{Q}(\mathbf{r}),
\end{equation*}
such that the unity operator is given by
\begin{equation*}
	\mathbf{1} = \sum_{iP\mu,jQ\nu} |iP\mu\rangle (\underline{S}^{-1})_{i\mu,j\nu}^{P,Q} \langle jQ\nu|.
\end{equation*}

The composition of several single crystals with a substitution level around $x=0.5$ was accurately determined by x-ray diffraction (XRD) using a STOE imaging plate diffraction system (IPDS-2T) equipped with Mo K$_\alpha$ radiation. All accessible symmetry-equivalent reflections were measured at RT up to a maximum angle $2\theta = 65$°. The data were corrected for Lorentz, polarization, extinction, and absorption effects. Using SHELXL \cite{Sheldrick} and JANA2006\cite{Petek} around $155$ averaged symmetry-independent reflections ($I > 2\sigma$) have been included for the respective refinements in space group \textit{P}4/\textit{nmm}. The refinements converged quite well and show excellent reliability factors (see \tableref{tab:table1}). The lattice parameters and atomic positions used in the electronic structure calculations were obtained from refinement of the XRD data. The lattice parameters for  an idealized $x=0.5$ crystal were the results of two FeSe$_{1-x}$Te$_{x}$ samples with $x=0.483$ and $x=0.516$, respectively. Crystallographic information regarding the refinement of both samples is listed in \tableref{tab:table1}. As DFT methods are notoriously inadequate for predicting structures of the pnictide family, relying on experimentally observed lattice parameters is well justified.

\begin{table}[t]
	\caption{\label{tab:table1} Structural parameters of FeSe$_{1-x}$Te$_{x}$ determined from single-crystal x-ray diffraction. The structure was refined in the tetragonal space group $P4/nmm$. Se/Te and interstitial Fe2 sit on $2c$ Wyckoff positions with coordinates $\frac{1}{2}, 0, z$ whereas Fe1 sits on a special position $2a$ with coordinates $0,0,0$.\@ The $U_{ii}$ denote the anisotropic atomic displacement parameters (for Fe1 and Se/Te $U_{11}=U_{22}$ and $U_{12}=U_{13}=U_{23}=0$); for interstitial Fe2 only $U_{\rm iso}$ is given. Refinement of the site occupancy factor (SOF) of Fe2 demonstrates that the  Se/Te substituted samples contain a significant amount of interstitial Fe.\@}
	\begin{ruledtabular}
		\begin{tabular}[t]{llccc}
			& FeSe$_{1-x}$Te$_{x}$ & $x=0$ & $x=0.483(9)$ & $x=0.516(8)$\\
			\hline
			&  $a$ (\r{A}) & 3.7688(7) & 3.7913(7) & 3.7948(2) \\
			&  $c$ (\r{A}) & 5.520(1) & 5.945(3) & 5.986(1)  \\
			&  & &  &\\
			Fe1        & $U_{11}$ (\r{A}$^2$) & 0.0108(5) & 0.0096(2) & 0.0106(1) \\
			& $U_{33}$ (\r{A}$^2$) & 0.0226(6) & 0.0184(5) & 0.0187(2) \\
			&  & & & \\
			Se/Te        & $z$ & 0.26680(9) & 0.27794(9) & 0.27984(7) \\
			& $U_{11}$ (\r{A}$^2$) & 0.0138(4) & 0.0126(1) & 0.0129(1) \\
			& $U_{33}$ (\r{A}$^2$) & 0.0184(4) & 0.0365(4) & 0.0368(2) \\
			&  & & &\\
			Fe2        & $z$ & $-$ &  0.6969(16) & 0.6991(9) \\
			& $U_{\rm iso}$ (\r{A}$^2$) & $-$ & 0.0111(21) & 0.0134(12) \\
			& SOF & $-$ & 0.080(4) &  0.105(3) \\
			&  & & &\\
			& $wR_2$ (\%) & 4.67 & 2.70 &  3.56 \\
			& $R_1$ (\%) & 1.92 & 1.49 &  1.53  \\
		\end{tabular}
	\end{ruledtabular}
\end{table}
 The Fe(Se,Te) crystals exhibit an interstitial site, in between the iron planes (see Fe2 in \tableref{tab:table1}). This interstitial iron has significant effects on the superconducting and magnetic properties of the system and has been subject of extensive research\cite{inter1,inter2,inter3}. This excess iron can lead to a suppression of superconductivity, which would render the system inadequate for the search for Majorana bound states. However, it was shown that superconductivity persists at low interstitial content\cite{inter2} and that excess iron may even be reduced from the as-grown samples and superconductivity enhanced via annealing\cite{Sun_2019}. Furthermore, substitution on the Fe site with transition metals, as considered in \secref{sec:electrondoping}, might affect the interstitial site as well. The questions that arise in the context of excess iron are, however, beyond the scope of this paper and will be investigated via the CPA-method in future works. Thus, for the following theoretical considerations we neglect the interstitial site. 

\section{Results and Discussion}\label{sec:Results}
\subsection{Bandstructure of FeSe}
\begin{figure}[tb]
	\includegraphics[width=\linewidth]{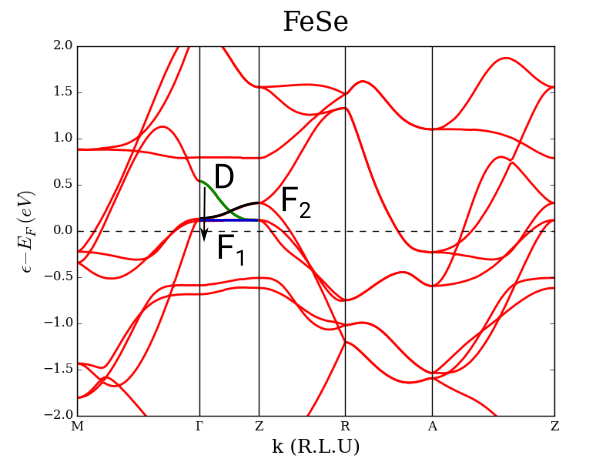}
	\caption{Bandstructure (red lines) of FeSe with lattice parameters $a=3.7688$\r{A}, $c=5.520$\r{A}, $z=0.2668$. Green line highlights dispersive $p_{z}$-character band labelled \textit{D}, blue line highlights nondegenerate flat $d$-character band labelled \textit{F$_1$} and black line highlights twofold degenerate flat $d$-character band labelled \textit{F$_2$}. The arrow indicates effect of Te-substitution.}
	\label{fig:FeSe}
\end{figure}

The band structure of FeSe has been studied in great detail by ARPES measurements \cite{Amir1,Amir2,Amir3}, but
we restrict our discussion to the $\Gamma$Z line. In \secref{sec:SOC}, we will include spin-orbit coupling (SOC), but for the time being we will neglect it. As can be seen from the DFT bandstructure in \figref{fig:FeSe}, the $\Gamma$Z line only shows minimally dispersive bands close to the Fermi energy, attributed to $3d$-Fe-orbitals. As a result, FeSe exhibits two-dimensional behaviour with \textit{intra}layer hopping but only minimal \textit{inter}layer hopping. The lower of these two bands ($d_{xy}$ orbitals, labelled \textit{F$_1$} in \figref{fig:FeSe}) is non-degenerate, while the upper band exhibits a two-fold degeneracy ($d_{xz}/d_{yz}$ orbitals, labelled \textit{F$_2$} in \figref{fig:FeSe}).
Located above the $F_1$ band is a highly dispersive  band with $p_z$-character (labelled \textit{D} in \figref{fig:FeSe}), that can be affected by the chalcogen, i.e., by Te-substitution. The goal is to induce a band inversion by lowering the \textit{D}, thus inverting the $p_z$- and $d_{xz}$/$d_{yz}$ bands at Z. Together with SOC, this will open up a gap at the crossing point with the F$_2$ band. This was shown to result in a topological bandstructure with a surface Dirac cone (SDC) that could host Majorana bound states in the superconducting phase\cite{ZhangPRB,ZhangScience,Wang333}. 

However, the Dirac cone in FeSe$_{0.5}$Te$_{0.5}$ is situated well above the Fermi-level and recent attempts to access it via surface deposition\cite{ZhangScience}, while confirming the SDC in ARPES measurements, inevitably distort surface transport. To preserve surface transport we consider co-substitution, i.e., intrinsic doping, in order to lower the SDC towards the Fermi-level. In addition, our application of the CPA will generate insight into the nature of disorder in these compounds which is beyond the DFT super-cell calculations of \refcite{ZhangPRB}. It gives access to information on level shifts and band broadening. With the considerable disorder induced via substitution, it is crucial to verify whether the involved quasi-particles remain well defined and the SOC gap unobstructed.

\subsection{Effect of Te-substitution}
\begin{figure}[tb]
	\includegraphics[width=\linewidth]{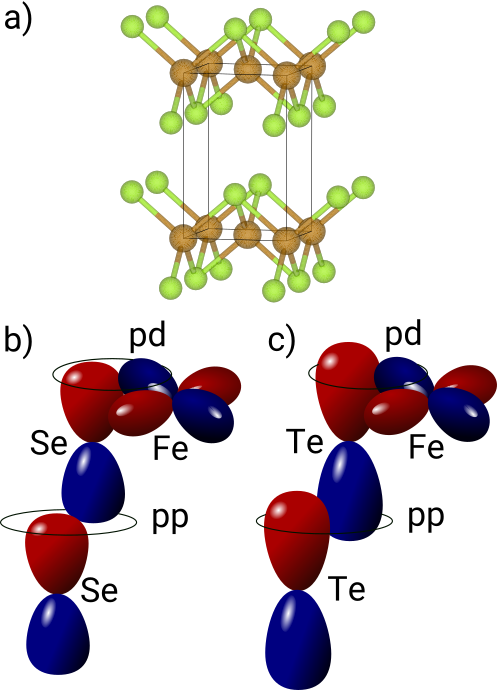}
	\caption{Schematic drawing of (a) Fe$X$ (X=\{Se,Se$_{0.5}$Te$_{0.5}$\}) structure (brown and green balls represent Fe and $X$ atoms, respectively) and overlap of Fe- and (b) Se-orbital, (c) Te-orbital between the iron planes. Schematic orbitals in the style of Fig. 1 of \refcite{ZhangPRB}.}
	\label{fig:orb}
\end{figure}

The effect of Te-substitution on the bandstructure of FeSe is most commonly attributed to the spatial extent of the $p_z$-orbitals of Te\cite{ZhangPRB}. Due to the limited overlap of $p_z$-orbitals of Se between Fe-layers and the resulting small hybridization along the $c$-axis, FeSe displays two-dimensional behaviour. Upon Te-substitution, the hybridization of $p_z$-orbitals between Fe-layers is increased, due to the greater spatial extent of Te orbitals (see schematics in \figref{fig:orb}(b)). 
\begin{figure}[tb]
	\centering
	\includegraphics[width=0.62\linewidth]{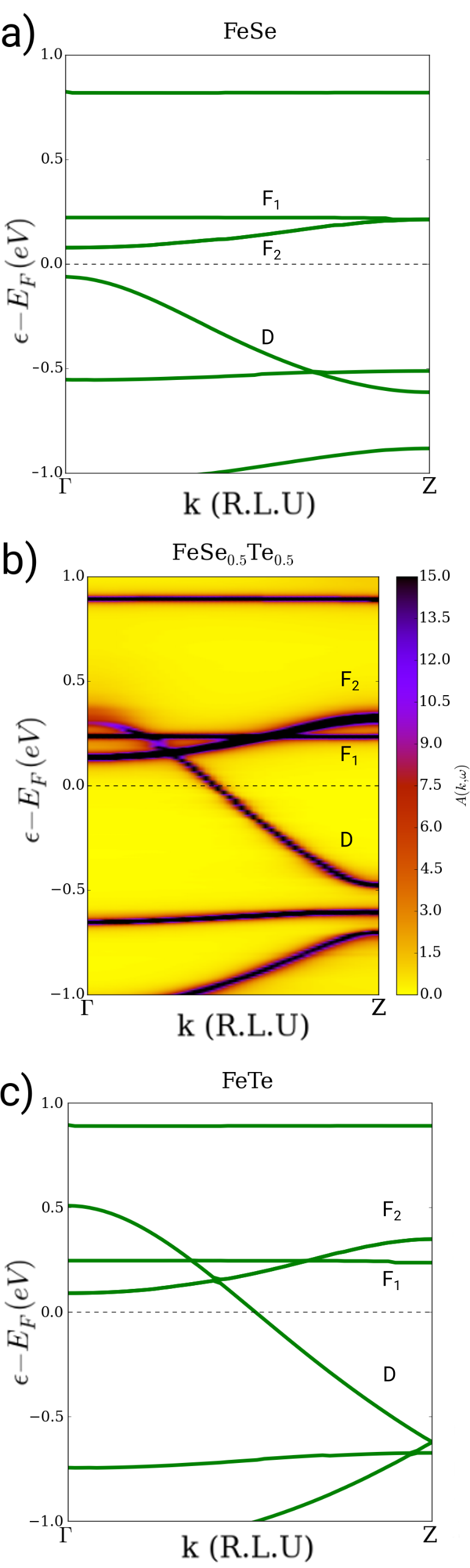}
	\caption{Along $\Gamma$Z (a) DFT bandstructure of FeSe, (b) CPA Bloch spectral function of FeSe$_{0.5}$Te$_{0.5}$ (c) DFT bandstructure of FeTe for lattice parameters $a=3.793$\r{A}, $c=5.9656$\r{A}, $z=0.27885$ of the substituted compound. Relevant bands labelled \textit{D}, \textit{F$_1$} and \textit{F$_2$}.}
	\label{fig:FeSeTe}
\end{figure}
\begin{figure}[tb]
	\includegraphics[width=\linewidth]{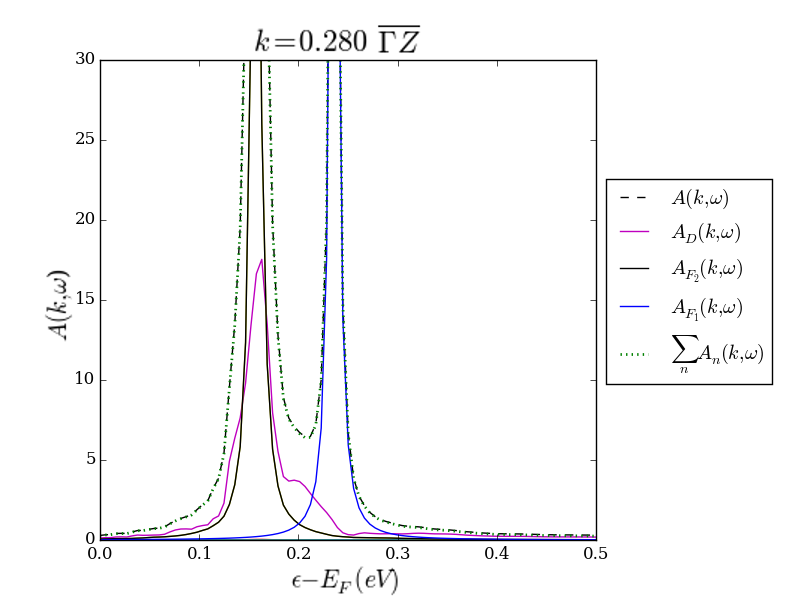}
	\caption{Bloch spectral function $A(\mathbf{k},\omega)$ of FeSe$_{0.5}$Te$_{0.5}$ along $\Gamma$Z at $\mathbf{k}=0.28 \overline{\Gamma \text{Z}}$, the lower band crossing point.}
	\label{fig:BD+A_FeSeTe}
\end{figure}
As a consequence, interlayer hopping is increased and the amplified $pp$-hybridization results in a highly dispersive $p_z$-character band in vicinity to the Fermi-level. In contrast to FeSe, Fe(Se,Te) exhibits three-dimensional behaviour.

This behaviour can be see from \figref{fig:FeSeTe}, in which we show the bandstructure of the pure end members FeSe and FeTe, and the Bloch spectral function of the substituted system FeSe$_{1-x}$Te$_{x}$ at $x=0.5$. 
with orbital overlap $\underline{S}$ and effective medium Green's function $\underline{\Gamma}$. Here, $\omega^+ = \omega + i\delta$ with infinitesimal $\delta$.
For comparison, we have adopted the lattice parameters of the substituted compound Fe(Se,Te) ($a=3.793$\r{A}, $c=5.9656$\r{A}, $z=0.27885$ from XRD) for both end members FeSe and FeTe. The bandstructure of this hypothetical FeSe crystal consequently differs from that in \figref{fig:FeSe}, calculated with real lattice parameters. While in the hypothetical FeSe the dispersive band \textit{D} is well below the Fermi-level and the $3d$ bands \textit{F$_1$} and \textit{F$_2$} (\figref{fig:FeSeTe}(a)), it crosses the flat bands in hypothetical FeTe (\figref{fig:FeSeTe}(c)). In \figref{fig:FeSeTe}(b) we show the Bloch spectral function $A(\mathbf{k},\omega)$ of FeSe$_{0.5}$Te$_{0.5}$ along the $\Gamma$Z line in false colour, which exhibits a behaviour intermediate between the clean compounds. The Bloch spectral function is calculated according to 
\begin{equation}\label{eq:A}
	A(\mathbf{k},\omega)= -\frac{1}{\pi}\Im \text{Tr} \bigl[ \underline{S}\underline{\Gamma}(\mathbf{k},\omega^+)\bigr],
\end{equation} 
The bandstructure of FeSe$_{0.5}$Te$_{0.5}$ was shown to be topologically non-trivial by \refscite{ZhangPRB,TopCharacter}, due to a band inversion. This is clearly visible in \figref{fig:FeSeTe}.
If we include perturbatively the spin-orbit interaction  a gap at the crossing point of \textit{D} and \textit{F$_2$} (which splits into \textit{F$_2^+$} and \textit{F$_2^-$} under SOC)  opens. This allows for a SDC that may host topologically protected surface states, especially Majorana bound states in a vortex in the superconducting phase\cite{PRL117}.

To get further insight into the effect of substitutional disorder, we disentangle the spectral peaks of the individual bands by projecting the $\mathbf{k}$-dependent Green's function $S(\mathbf{k})\Gamma(\mathbf{k},\omega)S(\mathbf{k})$ onto the eigenvectors of the clean parent compound. This is accomplished by defining a band-projected Green's function
\begin{equation}\label{eq:proj}
	G_n (\mathbf{k},\omega) \equiv \sum_{i,j\in \text{parent}} c_{n,i}^{*}(\mathbf{k})[S(\mathbf{k})\Gamma(\mathbf{k},\omega)S(\mathbf{k})]_{i,j}c_{n,j}(\mathbf{k}),
\end{equation}
with the $j$th orbital component $c_{n,j}$ of the eigenvector of band $n$. Because the eigenvectors are defined on the smaller Hilbert space of the parent compound, the sum runs only over orbital indices of that subspace. \figref{fig:BD+A_FeSeTe} shows the projected spectral function of FeSe$_{0.5}$Te$_{0.5}$ at the crossing point of band \textit{D} and \textit{F$_2$}. As can be seen, the crossing point lies well above the Fermi-level at $\epsilon\simeq 0.16$eV, thus making it inaccessible to experiment. We will address this problem in \secref{sec:electrondoping} and show how the crossing point can be shifted towards the Fermi-level.

\subsection{Electron doping via chemical substitution}\label{sec:electrondoping}
In order to bring the band crossing point (see \figref{fig:FeSeITe1}(a)) closer to the Fermi-level, we consider a co-substitution. By bringing additional charges into the bulk system, we circumvent the disruption of surface transport due to surface deposition, as proposed by \refcite{ZhangPRB}. To this end, we follow two different strategies: Firstly, a substitution of Fe by transition metals, namely Co, Cu and Ni, respectively. Secondly, a co-substitution on the Se site. 
\subsubsection{Fe site co-substitution}\label{sec:FeCosub}
\begin{figure}[tb]
   	\includegraphics[width=\linewidth]{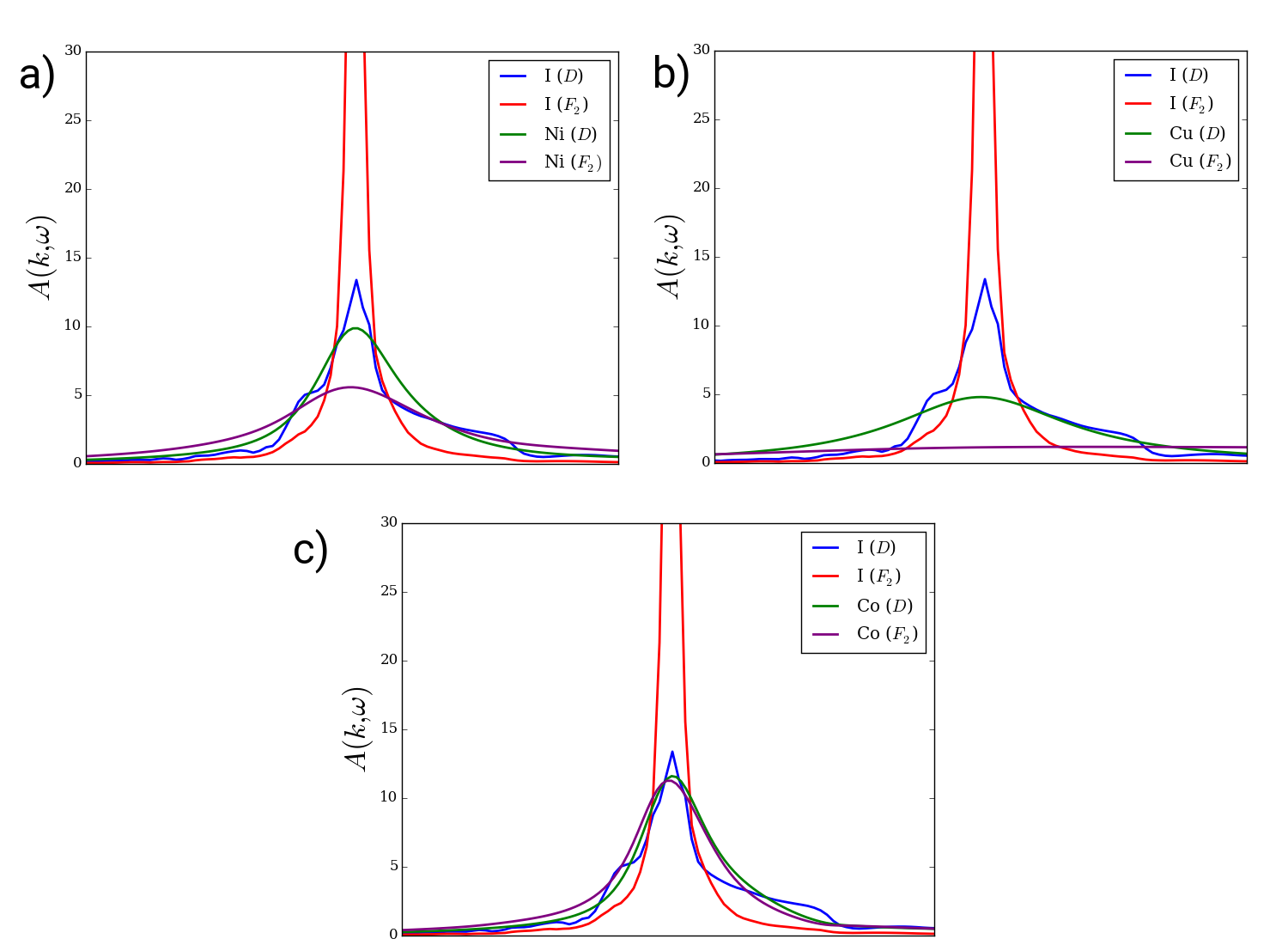}
   	\caption{Comparison of the projected Bloch spectral function $A(\mathbf{k},\omega)$ of FeSe$_{0.325}$I$_{0.175}$Te$_{0.5}$ for bands $D$ and $F_2$ at their respective crossing point to (a) Fe$_{0.85}$Ni$_{0.15}$Se$_{0.5}$Te$_{0.5}$, (b) Fe$_{0.75}$Cu$_{0.25}$Se$_{0.5}$Te$_{0.5}$ and (c) Fe$_{0.6}$Co$_{0.4}$Se$_{0.5}$Te$_{0.5}$. Peaks were shifted in energy to coincide for better comparison.}
   	\label{fig:CompNiCuCo}
\end{figure}

 While all three candidates did in fact raise the Fermi-level there are two major drawbacks which excluded this strategy: Firstly, for all three candidates, the necessary substitutional degree $y$ in Fe$_{1-y}$\textit{M}$_{y}$Se$_{0.5}$Te$_{0.5}$ (\textit{M} = Ni, Co, Cu) was relatively high ($y=0.15-0.4$), resulting in pronounced spectral broadening (see \figref{fig:FeSeITe1}). To emphasize this, in \figref{fig:CompNiCuCo} we compare the projected spectral function of the respective $D$ and $F_2$ bands to iodine co-substitution at the Se site, our most promising candidate (see discussion below). It is evident from this comparison that Co, Cu and Ni result in larger broadening of the bands connected to the SOC gap, thus increasing the probability of concealing said gap and prohibiting a SDC. The main drawback of the Fe site substitution, however, is  the fact that already at very low concentrations ($y\simeq 0.05$), superconductivity is suppressed due to the strong scattering properties of Ni, Co and Cu\cite{CoDoped1,CoDoped2,CoNiDoped}. This excludes such compounds from the search for Majorana bound states.
\begin{figure}[tb]
 	\includegraphics[width=\linewidth]{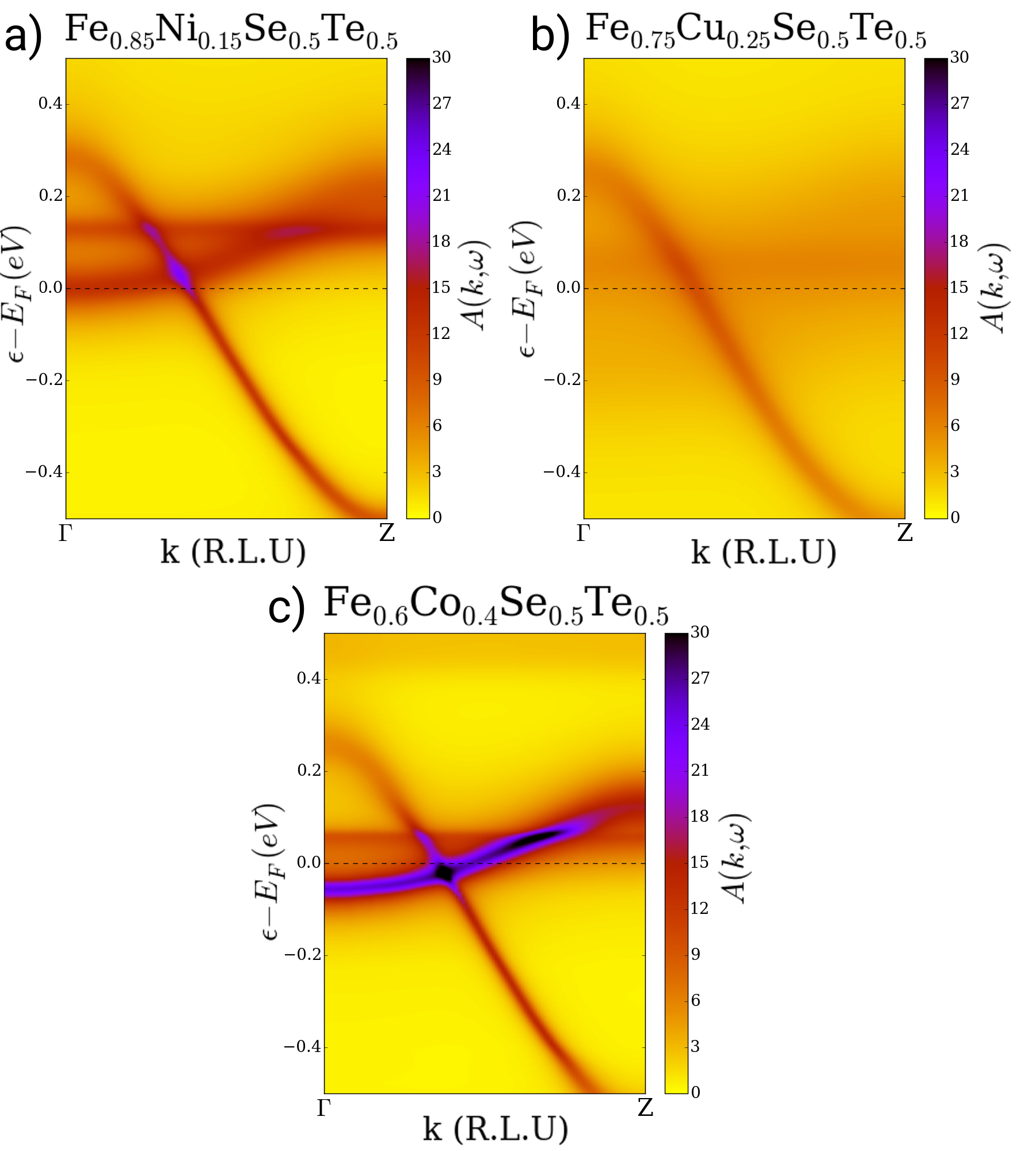}
 	\caption{Comparison of the Bloch spectral function $A(\mathbf{k},\omega)$ along $\Gamma$Z of (a) Fe$_{0.85}$Ni$_{0.15}$Se$_{0.5}$Te$_{0.5}$ (b) Fe$_{0.75}$Cu$_{0.25}$Se$_{0.5}$Te$_{0.5}$. and c) Fe$_{0.6}$Co$_{0.4}$Se$_{0.5}$Te$_{0.5}$.}
 	\label{fig:NiCuCo}
\end{figure}
\subsubsection{Se site co-substitution}\label{sec:SeCosub}
We have identified the most promising candidate for this co-substitution to be iodine. 
\begin{figure}[tb]
	\includegraphics[width=\linewidth]{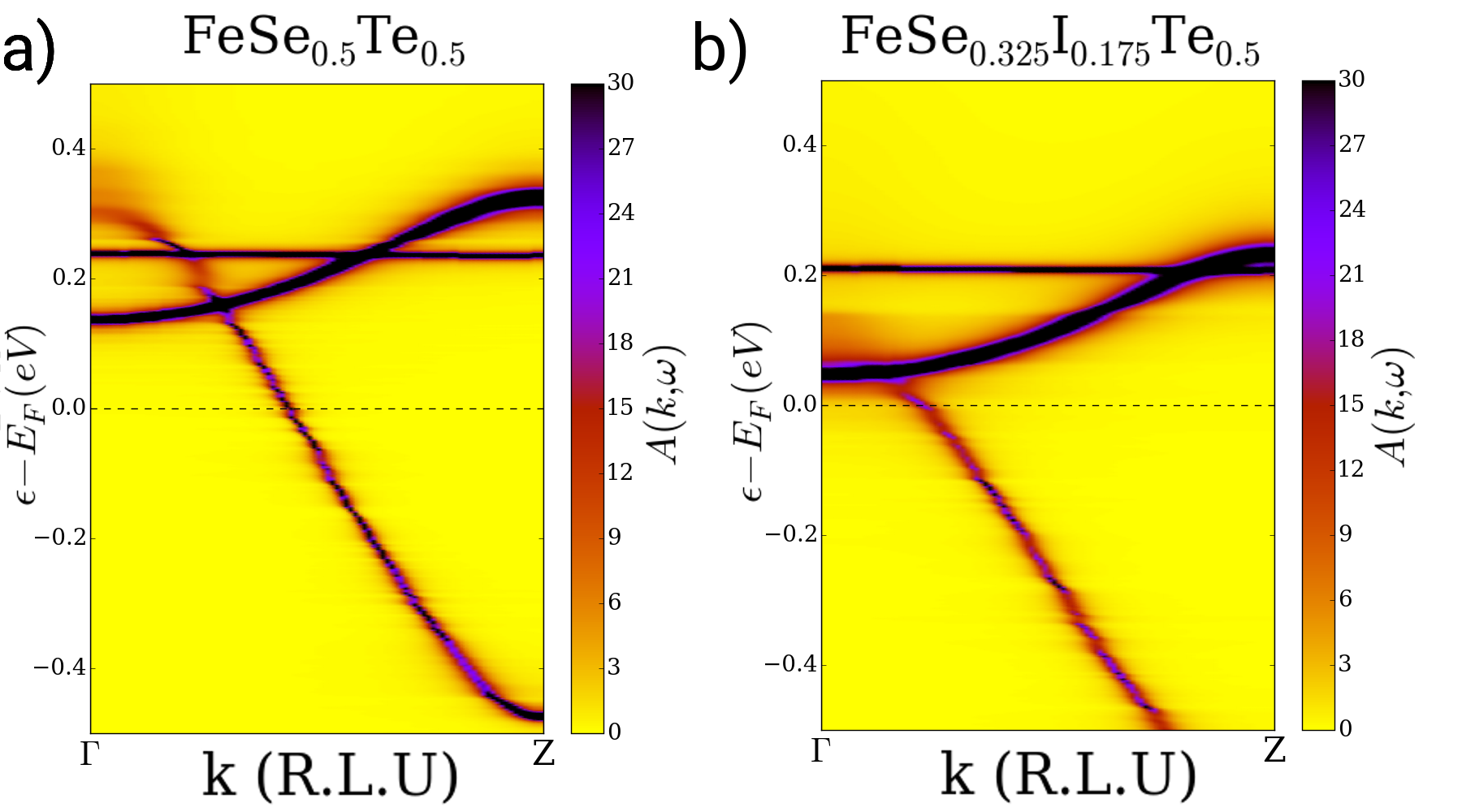}
	\caption{Comparison of the Bloch spectral function $A(\mathbf{k},\omega)$ along $\Gamma$Z of (a) FeSe$_{0.5}$Te$_{0.5}$ (b) FeSe$_{0.325}$I$_{0.175}$Te$_{0.5}$ without SOC.}
	\label{fig:FeSeITe1}
\end{figure}
Our choice has two distinct reasons: Firstly, iodine brings an additional valence electron into the bulk system, compared to Se and Te, thus raising the Fermi-level. 

Secondly, due to its close similarity to Te, especially with regard to ion-radii, we may expect iodine not to alter the crystal lattice parameters significantly. This is essential to our calculations, as they depend on the lattice parameters as input. Without a grown and fully characterized FeSe$_{1-x-y}$I$_{y}$Te$_{x}$ crystal, we must rely on parameters that are reasonable for the hypothetical crystal structure. To this end, we adopt the FeSe$_{0.5}$Te$_{0.5}$ parameters for the co-substituted calculations.

As is evident from the comparison of \figref{fig:FeSeITe1}(a) and (b), the co-substitution of iodine ($y=0.175$, $x=0.5$) raises the Fermi-level, bringing the band crossing point from $\epsilon\simeq 0.16$eV down to $\epsilon\simeq0.05$eV (without SOC). It is at this crossing point, that SOC opens up a gap in which the surface Dirac cone resides, which now becomes experimentally accessible. 

Aside from iodine substitution at the Se site we further considered Br as a possible candidate. Our investigations into Br co-substitution showed that the desired effect of raising the Fermi-level could be achieved. However, at the same time the dispersive band $D$ was lowered below the degenerate $F_2$ band, thus not resulting in a band crossing and rendering Br inadequate. This leaves iodine as the only viable candidate. However, both candidates could serve as fine tuning parameters for the manipulation of the considered bands.

\subsection{Effect of spin-orbit coupling}\label{sec:SOC}
\begin{figure}[tb]
	\includegraphics[width=\linewidth]{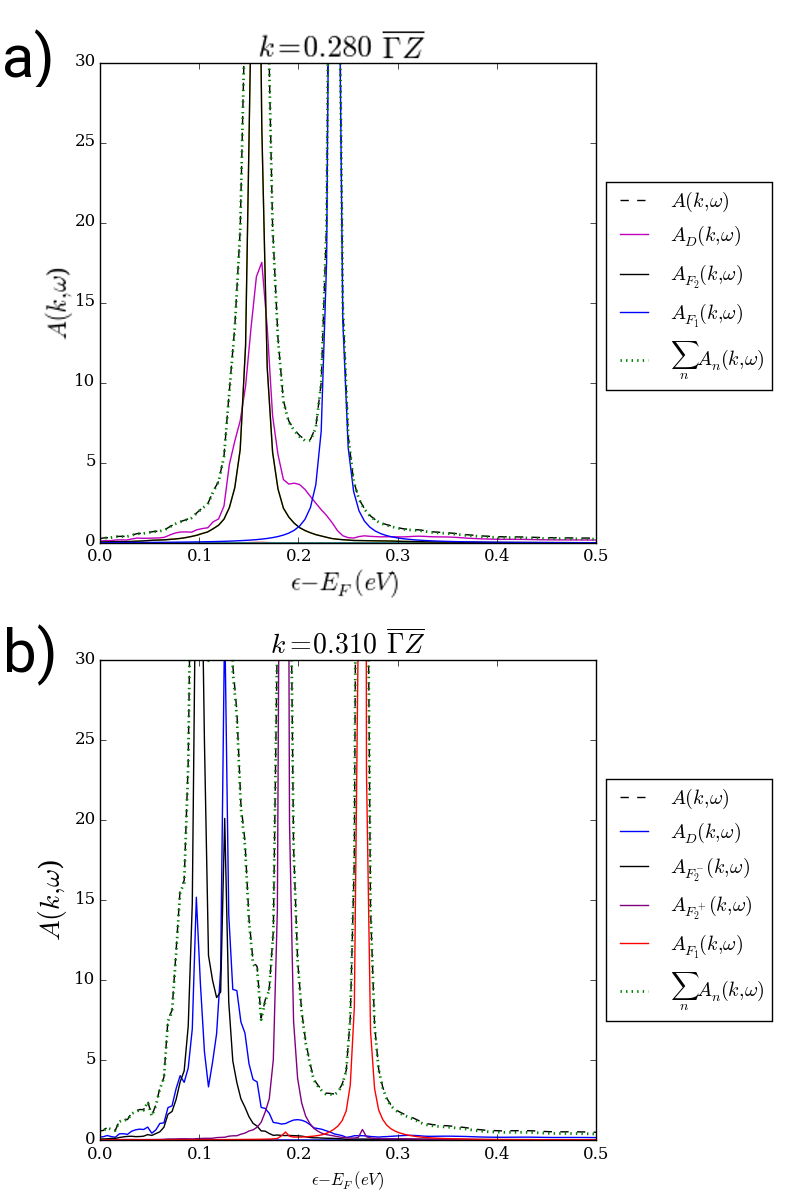}
	\caption{Bloch spectral function of FeSe$_{0.5}$Te$_{0.5}$ along $\Gamma\text{Z}$ at band crossing point (a) $\mathbf{k}=0.28 \overline{\Gamma\text{Z}}$ without SOC and at (b) $\mathbf{k}=0.31 \overline{\Gamma\text{Z}}$ with SOC.}
	\label{fig:BlochSOC}
\end{figure}

\begin{figure}[tb]
	\includegraphics[width=\linewidth]{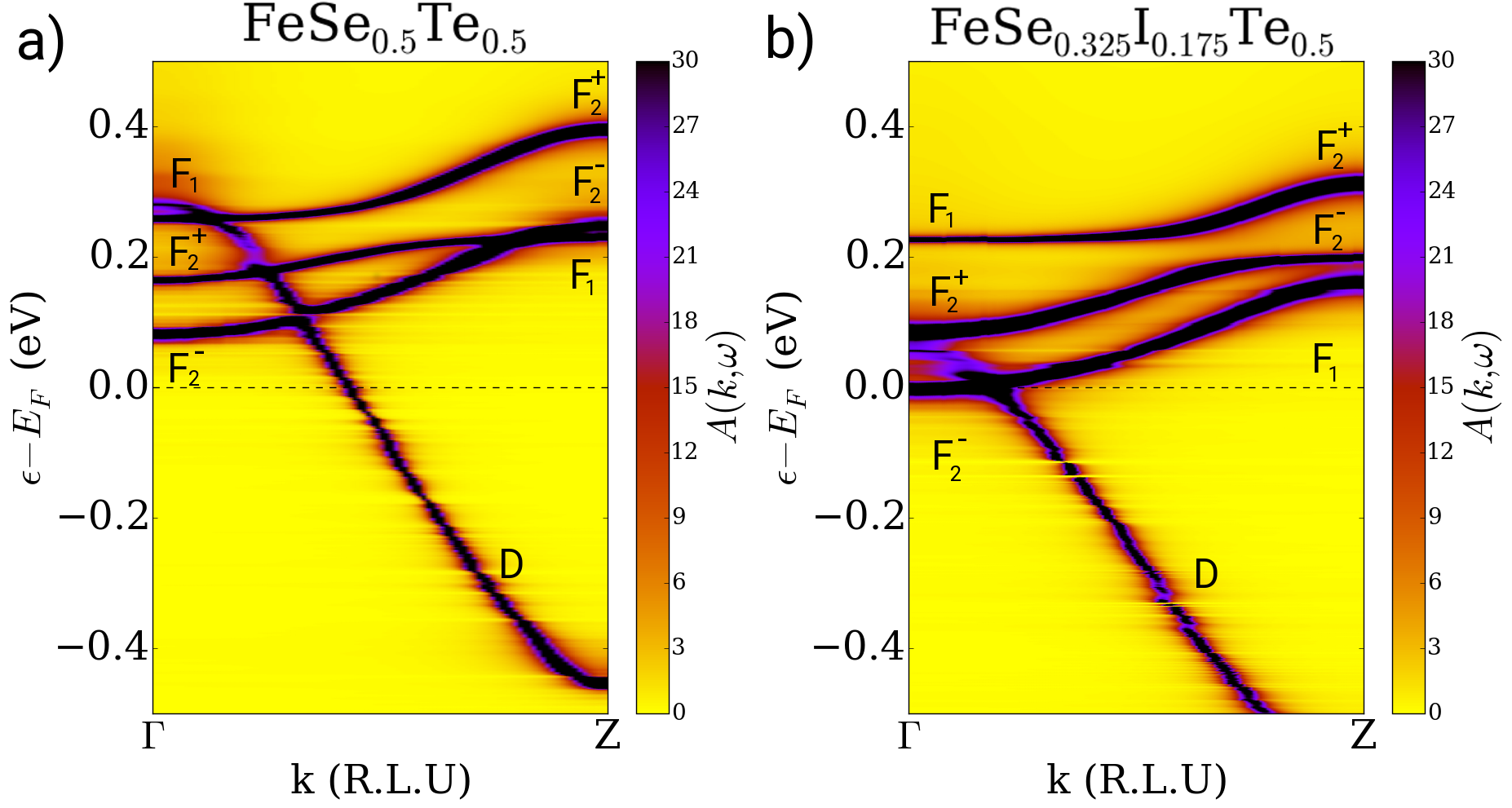}
	\caption{Comparison of the Bloch spectral function $A(\mathbf{k},\omega)$ along $\Gamma$Z of (a) FeSe$_{0.5}$Te$_{0.5}$ with SOC (b) FeSe$_{0.325}$I$_{0.175}$Te$_{0.5}$ with SOC.}
	\label{fig:FeSeITe2}
\end{figure}

\begin{figure}[tb]
	\includegraphics[width=\linewidth]{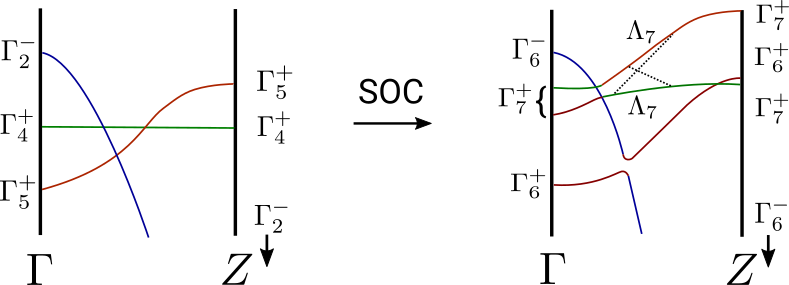}
	\caption{Schematics of the effect of SOC on the $\Gamma$Z line of FeSe$_{0.5}$Te$_{0.5}$. Figurative states at $\Gamma$ and Z are labelled according to their irreducible representations and parities (see \refcite{ZhangPRB}).}
	\label{fig:irred}
\end{figure}

We now turn to the SOC gap in which the SDC appears and to the question whether it survives the additional iodine substitution proposed in \secref{sec:electrondoping}.
The results presented in this section were obtained by a full \textit{ab-initio} treatment of SOC within the CPA (see \appref{app:SOC}). The main effect of SOC in Fe(Se,Te) is lifting the degeneracy of band $F_2$, splitting into $F_2^{+}$ and $F_2^-$ (see \figref{fig:FeSeITe2}(a)), and opening up a SOC gap of $\Delta_{\text{SO}} \simeq 30$meV at the crossing point ($\epsilon\simeq 0.109$eV) of band $D$ and $F_2^-$. As $F_2^-$ is shifted downward in energy relative to $D$, the crossing points of the system \textit{without} and \textit{with} SOC no longer coincide in $\vec{k}$-space. This explains why the comparison in \figref{fig:BlochSOC} is made at different $\vec{k}$-points along $\Gamma$Z.
The opening of the gap becomes evident from examining the projected spectral function in \figref{fig:BlochSOC}. Clearly, bands $D$ and $F_2^-$ split into two distinct peaks each and transfer spectral weight across the gap.  Consequently, a Dirac cone that can host topologically non-trivial surface states forms on the surface, within the SOC gap. The gap size is in good agreement with ARPES measurements of \refcite{ZhangScience}. Additionally, some minor mixing of bands $F_2^+$ and $F_1$ may be observed from the respective projected spectral functions. This mixing increases as we approach the high symmetry point Z and can be explained by looking at the irreducible representations (of the point-group $D_{4h}$) of states at $\Gamma$ and Z connected by these bands (see schematics in \figref{fig:irred}). 
Here, we follow the nomenclature of \refcite{ZhangPRB}. Without SOC, the states of bands $F_2$ at $\Gamma$ transform as $\Gamma_{5}^{+}$, while the state of band \textit{F$_1$} transforms as $\Gamma_{4}^{+}$. With SOC, the former bands split and their states now transform as $\Gamma_{6}^{+}$ and $\Gamma_{7}^{+}$, respectively. The state of band F$_1$ at $\Gamma$ now transforms as $\Gamma_{7}^{+}$ and the states of $F_2^+$ and F$_1$ along the high symmetry line $\Gamma\text{Z}$ both transform as $\Lambda_{7}$. Due to their similar character and close proximity, they strongly mix.
\begin{figure}[tb]
	\includegraphics[width=0.9\linewidth]{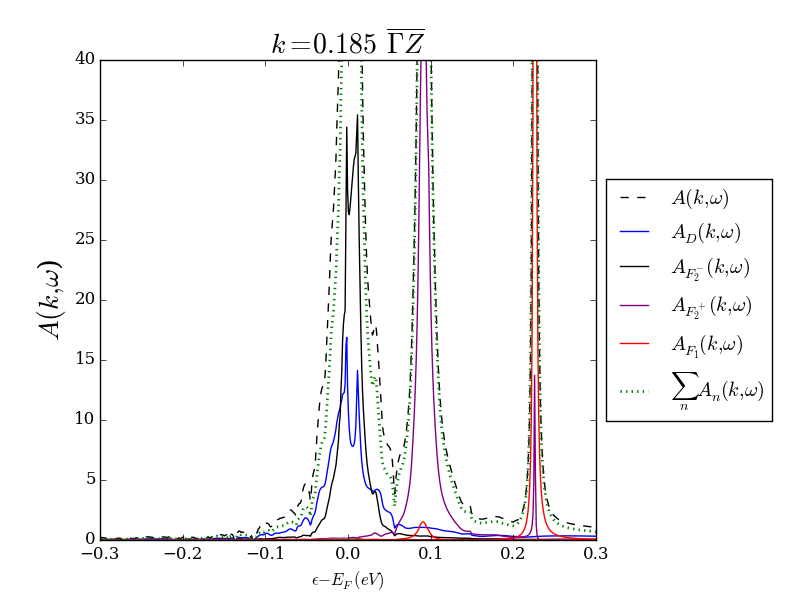}
	\caption{Bloch spectral function $A(\mathbf{k},\omega)$ of FeSe$_{0.325}$I$_{0.175}$Te$_{0.5}$ with SOC at the band crossing point $\mathbf{k}=0.185\overline{\Gamma\text{Z}}$.}
	\label{fig:FeSeITeBlochSOC}
\end{figure}
Similar effects are observed for Fe(Se,Te,I) with SOC, whose bandstructure we present in \figref{fig:FeSeITe2}(b). As in Fe(Se,Te) we observe clear band splitting due to SOC and a pronounced mixing of bands $F_1$ and $F_2^+$. Though indiscernible in \figref{fig:FeSeITe2}(b), the gap in FeSe$_{0.325}$I$_{0.175}$Te$_{0.5}$ becomes clear from examining the projected spectral function with SOC. 
The quasi-particle peaks of band $D$ and $F_2^-$ split up, shifting spectral weight across the gap. Clearly, the SOC gap ($\Delta_{\text{SO}}\simeq 10$ meV) survives the co-substitution and now the gap is centred at $\epsilon\simeq 5$ meV, closer to the Fermi-level at.
\section{Conclusion}\label{sec:Conclusion}
In this paper, we have studied the effect of intrinsic doping on the position of the surface Dirac cone of the Fe-based superconductor FeSe$_{1-x}$Te$_{x}$ using the coherent potential approximation. We have shown that, by band-engineering this compound via intrinsic doping, the band crossing point in the $\Gamma\text{Z}$-line of Fe(Se,Te), which is crucial to the non-trivial topology and surface Dirac cone, can be brought down to the Fermi-level. Apart from the successful iodine co-substitution at the Se site, we were able to exclude further candidates (Br) and co-substitutions at the Fe site (Co, Cu, Ni). Our calculations show the survival of the SOC gap in the co-substituted system, suggesting a stable surface Dirac cone and stable surface states. Thus,  we find FeSe$_{1-x-y}$Te$_{x}$I$_{y}$ ($x=0.5$, $y=0.175$) to be a promising candidate for a topologically non-trivial, single crystal superconductor that may host Majorana bound states.
\acknowledgments
This work was supported by the Virtual Materials Design initiative within the Helmholtz program Science and Technology of Nanosystems at the Karlsruhe Institute of Technology. The authors thank Peter Schweiß for fruitful discussions and acknowledge support by the state of Baden-Württemberg through bwHPC. The contribution from M.M. was supported by the Karlsruhe Nano Micro Facility (KNMF). All plot within this paper were generated by matplotlib\cite{MatPlot}, an open source project.

\appendix

\section{The  {\em ab initio} version of the coherent potential approximation due to Blackman, Esterling and Berk} 
In this appendix we summarize the main steps of the {\em ab initio} version of the coherent potential approximation (CPA) due to Blackman, Esterling and Berk\cite{BEB}. We start our discussion with a brief summary of the CPA approach applied to a single particle Hamiltonian with random on-site energies\cite{PhysRev.156.1017,PhysRev.156.809}.

\subsection{A brief review of the conventional CPA}
We briefly review the conventional CPA and refer the interested reader to \refcite{Herbig1} for a more detailed discussion. The most convenient starting point for the description of a substitutionally disordered crystal in a localized framework is a single-particle Hamiltonian of the form
\begin{equation}\label{eq:Ham}
\hat{H} = \sum_{i,j} W_{i,j}c_{i}^{\dagger}c_{j} + \sum_{i} \epsilon_{i}c_{i}^{\dagger}c_{i}.
\end{equation}
Here, $c_{i}^{\dagger} (c_{i})$ represent fermionic creation (annihilation) operators, $W_{i,j}$ denotes the hopping element of an electron between sites $i$ and $j$, and $\epsilon_{i}$ is a randomly-distributed onsite energy.
The substitutional disorder of the model Hamiltonian in \equref{eq:Ham} enters via the onsite terms, i.e., one assumes random onsite energies $\epsilon_{i}$. In such a scenario the distribution of energy levels of a given  site is included in the CPA, yet disorder at surrounding sites are only treated on average, i.e., correlations of a given site with the disorder of its environment  are neglected.

Within the CPA\cite{PhysRev.156.1017,PhysRev.156.809}, the disordered crystal is replaced by an effective medium associated with an effective medium Green's function $\Gamma$ and a self-energy $\Sigma$. The assumption, consistent with the mentioned neglect of inter-site correlations, that may be imposed upon this self-energy is to take it as a single-site quantity. In this sense the CPA is the dynamical mean field theory of substitutionally disordered systems. Having established the effective medium, one may now replace a site of the medium with a real impurity with a well defined onsite energy. Due to the single-site nature of this theory, only the diagonal elements of the impurity Green's function $^q G$ of such an insertion are relevant and my be expressed as
\begin{equation}\label{eq:impG}
^q G_{i,i} = \bigl(\Gamma_{i,i}^{-1} + \Sigma_{i} - \epsilon_{i}^{q} \bigr)^{-1},
\end{equation}
where $q$ is the species index of the impurity and $i$ denotes a site. This replacement is then repeated with all species allowed at this site and it is demanded that these replacements must not change the effective medium on the average.

This postulated self-consistency condition can now be formulated as
\begin{equation}\label{eq:Gamma1}
	\Gamma_{i,i} \mbeq  \sum_{q} c_{i}^{q} \, ^q G_{i,i},
\end{equation}
with the atomic concentration $c_{i}^{q}$ of species $q$ at site $i$. In order for the effective medium Green's function to have physical meaning, it must coincide with the configurationally averaged Green's function $\langle G \rangle$ of the disordered system
\begin{equation}\label{eq:Gamma2}
	\Gamma_{i,i} \mbeq \langle G \rangle_{i,i} = \bigl[ \bigl(G^{0} \bigr)^{-1} - \Sigma \bigr]_{i,i}^{-1} = \bigl[\omega- W - \Sigma \bigr]_{i,i}^{-1}.
\end{equation}
In a periodic system, the configurationally averaged Green's function retains its full translational invariance and the self-energy is site independent, such that we may express $\Gamma$ in Fourier space as
\begin{equation}\label{eq:Gamma3}
	\Gamma_{i,i} = \int_{1.BZ} d^3 k \bigl[ \omega - W(\mathbf{k})-\Sigma \bigr]^{-1},
\end{equation}
with $W(\mathbf{k}) = \sum_{j} W_{0j}e^{i\mathbf{kR}_{j}}$, assuming one site per unit cell, for simplicity.
The set of self-consistent Eqs. \eqref{eq:impG}, \eqref{eq:Gamma1}, and \eqref{eq:Gamma2} or \eqref{eq:Gamma3} must now be solved in an iterative scheme.

The single-site nature of this method is one of its essential advantages, making it computationally feasible and thus generally applicable within \textit{ab initio} approaches. This comes at the expense of off-diagonal disorder: the hopping matrix elements lack the influence of the disordered environment surrounding a particular site.
\subsection{The Blackman, Esterling and Berk formalism}
An approach to improving the CPA method was published by Blackman, Esterling and Berk (BEB)\cite{BEB}. Here, we briefly outline the most important statements and refer the interested reader to \refcite{Herbig1} for a detailed account. Blackman, Esterling and Berk introduced occupation variables
\begin{equation}
	\eta_{i}^{P} = \Biggl\{ \begin{matrix}
	1 & \text{if site $i$ is occupied with species $P$,}\\ 0 & \text{otherwise.}
	\end{matrix}
\end{equation}
The $\eta_i^P$ must obey the following set of conditions:
\begin{itemize}\label{it:BEBrules}
	\item[1.] Avoidance of multiple occupancy of a site by different species:\\ 
		\centerline{$\eta_{i}^{P}\eta_{i}^{Q} = \delta_{PQ}\eta_{i}^{P}$}
	\item[2.] Forced occupation of each site by exactly one species: \\
		\centerline{$\sum_{P} \eta_{i}^{P}=1$}
	\item[3.] Association of the configurational average of random variables with atomic concentrations $c_{i}^{P}$ of species $P$ at site $i$:\\
	\centerline{$\langle \eta_{i}^{P}\rangle = c_{i}^{P}$}
\end{itemize}

With these variables we can project a non-stochastic extended Hilbert space containing all configurations (underlined symbols) to a specific configuration in a reduced Hilbert space (normal symbols). Accordingly, the Hamiltonian of the BEB-CPA can be expressed as
\begin{align}\label{eq:BEBHam}
	\hat{H} &= \sum_{i,j,P,Q} \underline{H}_{i,j}^{P,Q}\eta_{i}^{P}\eta_{j}^{Q}c_{i}^{\dagger}c_{j} \\
	&= \sum_{i,j,P,Q} \underline{W}_{i,j}^{P,Q}\eta_{i}^{P}\eta_{j}^{Q}c_{i}^{\dagger}c_{j} + \sum_{i,P} \underline{\epsilon}_{i}^{P}\eta_{i}^{P}c_{i}^{\dagger}c_{i}.
\end{align}
With $\eta$ being the only stochastic quantities of the formalism, one can now select the Hamiltonian of a specific configuration. The non-stochastic quantities in the extended Hilbert space, $\underline{H}$, $\underline{W}$ and $\underline{\epsilon}$, posses the full translational symmetry of the clean crystal, making an implementation within an \textit{ab initio} scheme highly convenient. A further advantage of this  formalism is the inclusion of environmental disorder effects on the hopping elements $W_{i,j}$, which, in addition to the onsite terms $\epsilon_{i}$, now become random due to the set of $\eta$.

Under the BEB-transformation the Green's function in extended Hilbert space reads\cite{GonisBEB,Brouers}
\begin{equation}
	\underline{G}_{ij}^{PQ} = \eta_{i}^{P}G_{ij}\eta_{j}^{Q}.
\end{equation}
While only being a simple number in the conventional CPA, the site matrix element $G_{ij}$ now becomes a matrix in species space and the equations of motion become matrix equations.
We may now again \textit{define} a (BEB) self-energy $\underline{\Sigma}$ and effective medium Green's function $\underline{\Gamma}$.

Analogous to the conventional CPA, we introduce an impurity Green's function $^{q} \underline{G}$, which describes the insertion of a species $q$ at site $i$ of the effective medium:
\begin{equation}\label{eq:self1}
	^q \underline{G}_{ii}^{PQ} = \delta_{PQ}\delta_{Pq} \bigl[ [\underline{\Gamma}_{ii}^{-1}]^{qq} + [\underline{\Sigma}_{i}]^{qq}-\epsilon_{i}^{q} \bigr]^{-1}.
\end{equation}
In this extended formalism, the CPA self-consistency condition now becomes
\begin{equation}\label{eq:self2}
	\underline{\Gamma}_{ii}^{PQ} = \sum_{q} c_{i}^{q}\, ^q \underline{G}_{ii}^{PQ}
\end{equation}
and for a periodic system with a single site per unit cell, it again holds that
\begin{equation}\label{eq:self3}
	\underline{\Gamma}_{ii} = \int_{1.BZ} d^3 k \bigl[ \omega \underline{\mathbf{1}}-\underline{W}(\mathbf{k}) - \underline{\Sigma} \bigr]^{-1}.
\end{equation}
Eqs. \eqref{eq:self1}, \eqref{eq:self2} and \eqref{eq:self3} can be solved iteratively in analogy to the conventional CPA\cite{Rowlands_2009}.
\subsection{Spin-orbit coupling within the CPA}\label{app:SOC}
In order to incorporate SOC in form of a full \textit{ab-initio} treatment into the CPA, the formalism was extended to a spinor formalism. The spin-orbit coupling potential operator is then constructed from relativistic norm-conserving pseudo-potentials, as described in \refcite{RolfPRB}. Its real space representation is then given by a concentration weighted sum of contributions from all present atomic types:
\begin{equation}
	\langle \vec{r} | \hat{V}^{\text{SO}} | \vec{r}' \rangle = \sum_{i}c^q_{i} v_{q}^{\text{SO}}(\vec{r}-\vec{R_i},\vec{r}'-\vec{R_i}),
\end{equation}
where $c^q_i$ denotes the atomic concentration of atomic type $i$ and atomic site coordinate $\vec{R_i}$. For each atomic type $i$
\begin{equation}
	v_{q}^{\text{SO}}(\vec{r},\vec{r}') = \sum_{lm} \frac{\delta(r-r')}{r^2} v_{q,l}^{\text{SO}}(r) \mathbf{L_r}\cdot\mathbf{S}K_{lm}(\hat{r})K_{lm}(\hat{r}'),
\end{equation}
with cubic harmonics $K_{lm}$ and angular momentum and spin operators $\mathbf{L_r}$ and $\mathbf{S}$, respectively. Here, $\mathbf{L_r}$ acts on the $r$-coordinate. The angular momentum decomposed $v_{q,l}^{\text{SO}}$ are radial functions for each species $q$ and are given by the spin-orbit components of the norm-conserving relativistic pseudo-potential\cite{RolfPRB}. Due to the short-ranged nature of the potentials, the evaluation of their respective matrix elements is straight forward in the LCAO basis employed for the CPA and is done analogously to the evaluation of parts of the pseudo-potential (see \refcite{Herbig1} for details).

\bibliographystyle{apsrev4-1}
\bibliography{bande}
\end{document}